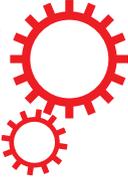

# OPEN

# Graphene-based extremely wide-angle tunable metamaterial absorber

Jacob Linder[1] & Klaus Halterman[2]

We investigate the absorption properties of graphene-based anisotropic metamaterial structures where the metamaterial layer possesses an electromagnetic response corresponding to a near-zero permittivity. We find that through analytical and numerical studies, near perfect absorption arises over an unusually broad range of beam incidence angles. Due to the presence of graphene, the absorption is tunable via a gate voltage, providing dynamic control of the energy transmission. We show that this strongly enhanced absorption arises due to a coupling between light and a fast wave-mode propagating along the graphene/metamaterial hybrid.

With the refinement and advancement of fabrication techniques and theoretical methods, it is now possible to successfully create anisotropic metamaterial structures that have at least one vanishing principal component of the permeability $\mu$ or permittivity $\varepsilon$ tensors. These artificial hybrid structures have been shown to allow propagation of electromagnetic (EM) waves through narrow[1] or mismatched[2] channels with little reflection, and offer alternatives in wavefront shaping[3,4], beam steering, and EM energy control[5] previously unseen in conventional systems. Typical experimental platforms involve various geometrical arrays of metallic rods[6,7] or combinations of metallic and dielectric multilayers that have tunable system parameters to achieve the desired near-zero effective EM response[8]. In addition to these platforms, other metamaterial architectures have been designed that manipulate the EM response to significantly reduce loss. These include, layered metal/dielectric structures[9,10], transparent conducting materials[11–13], superconducting nanoparticles[14,15], and plasmonic nanoshells with quantum dot cores[16]. While there are a number of mechanisms to dynamically tune metamaterials[17], it would be preferable to have more robust, real-time tunability of their effective electric ($E$) or magnetic ($H$) responses.

These developments have led to a significant number of recent works devoted to studying graphene as a component in hybrid metamaterial designs[18–48]. Enhanced absorbance in a hybrid graphene-metamaterial structure on a reflective substrate was proposed in ref. 26. In addition to its intrinsic 2D geometrical properties, graphene affords extensive control of its conductivity $\sigma$ due to its gate-tunable Fermi energy $E_F$. When EM waves interact with a graphene sheet, the gate voltage can tune the amount of energy that gets absorbed, depending on the relevant scattering processes, temperature, relaxation time, frequency of the incident beam and its polarization.

By placing graphene in contact with an anisotropic metamaterial that has vanishing principal components of the permittivity and permeability tensors over the frequencies of interest, one could expect the interplay between the two samples to present additional possibilities regarding the transmission of the light. More specifically, we pose the question if it is possible to utilize the dynamic and tunable conductivity of graphene to yield greater control over the way energy of EM waves is absorbed or reflected compared to when the materials are in isolation. Motivated by recent experimental and theoretical developments, we seek to establish if one may obtain tunable wide angle absorption in hybrid graphene/metamaterial layers where the tensor $\varepsilon$ describing the anisotropic response has components that are near zero[1].

To this end, we consider here the absorption characteristics of a layered system, comprised of monolayer graphene overlaying a metamaterial slab that is connected to a perfectly conducting metal. In accordance with the discussion above, we will consider metamaterial structures that can possess an $\varepsilon$ near-zero (ENZ) response, where the metamaterial has a vanishingly small value of the real part of $\varepsilon$. In contrast to numerous past works in the RF and visible light regime, we focus here on operational wavelengths that correspond to the less exploited THz regime, where many new technologies are emerging. To accurately determine the absorption as a function of the geometrical and material parameters that describe graphene and the metamaterial, Maxwell's equations are solved using a transfer matrix approach that accounts for the presence of graphene through boundary conditions imposed on the magnetic field. We discuss the behavior of the absorption in terms of the energy flow throughout the entire structure.

[1]Department of Physics, NTNU, Norwegian University of Science and Technology, N-7491 Trondheim, Norway. [2]Michelson Lab, Physics Division, Naval Air Warfare Center, China Lake, California 93555. Correspondence and requests for materials should be addressed to K.H. (email: klaus.halterman@navy.mil)

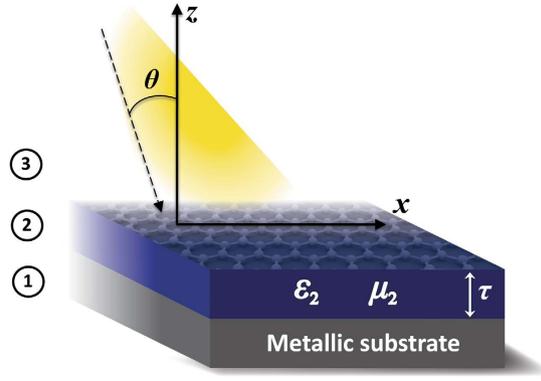

**Figure 1. Considered setup in this work.** An EM wave in vacuum (region 3) is incident upon a graphene layer coating an anisotropic ENZ metamaterial (region 2). A perfectly conducting metal backs the entire structure (region 1).

Our main result is that by tuning $E_F$, we are able to determine experimentally relevant regimes where *near perfect absorption can occur over an extremely wide range of incident angles for an ENZ metamaterial layer that has a small component of the permittivity tensor*. We show that the enhanced absorption occurs due to coherent perfect absorption which couples light to a fast wave propagating along the graphene/ENZ hybrid.

## Theory

The principal coordinate axes are chosen so that both the permittivity and permeability tensor in the anisotropic medium are both diagonal. A diagram of our setup is shown in Fig. 1. For concreteness, we focus on *p*-polarized incident light ($E_x, E_z, B_y$) ≠ 0. In the anisotropic medium (region 2), the nonzero components to the tensors are ($\varepsilon_{x,2}, \varepsilon_{z,2}, \mu_{y,2}$). It is backed by a perfect electric conductor (PEC) in region 1, where no electric or magnetic fields exist. The finite surface conductivity σ of graphene will be effectively taken into account in the boundary condition for the magnetic field *H* at the interface between air (region 3) and the anisotropic medium, where[26],

$$\sigma = 2ie^2 k_B T/(\pi \hbar^2)(\omega + i/t_{rel})^{-1} \ln[2 \cosh(E_F/(2k_B T))]. \quad (1)$$

When $E_F \gg k_B T, \hbar\omega$, we have,

$$\sigma = ie^2 E_F/[\pi \hbar^2 (\omega + i/t_{rel})]. \quad (2)$$

Thus, at a temperature $T$ of 300 K, and THz frequencies, we have $k_B T = 25$ meV, and $\hbar\omega \sim 1$ meV, meaning doping levels should exceed $E_F = 100$ meV when using Eq. (2). Here, $E_F$ is the Fermi energy, ω is the frequency of the EM wave, and $t_{rel}$ is the relaxation time. For frequencies in the THz region and lower, Eq. (2) reflects that the interband part to the conductivity is negligible compared its the intraband part, and consequently to the Drude-like behavior for σ.

We now consider Maxwell's equation (in SI units) in the incident air region and in the anisotropic medium, both having no free charges or currents. For *p*-polarized incident light and diagonal tensors, we obtain the following set of equations determining the components of the fields when the time-dependence is harmonic ($e^{-i\omega t}$):

$$\varepsilon_x \partial_x E_x + \varepsilon_z \partial_z E_z = 0, \quad i\omega B_y = -\partial_x E_z + \partial_z E_x, \quad -i\omega D_x = -\partial_z H_y, \quad -i\omega D_z = \partial_x H_y. \quad (3)$$

These equations are solved with a plane-wave ansatz for the field components where $\mathbf{E} \propto e^{ik_x x}$ and $k_x$ is conserved due to translational invariance. The Methods section contains additional details of the numerical and analytical solutions of the above equations.

Once the equations have been solved, one may compute the absorption $\mathcal{A}$ from the scattering coefficients: $A_2$ and $B_2$ are associated with the transmitted and reflected electric field wave in region 2 whereas $B_3 \equiv r$ is the coefficient for the reflected wave in region 3. Since there is no transmission into the PEC layer, the absorbance $\mathcal{A}$ can be computed from the incident wave after subtracting the reflected part: $\mathcal{A} = 1 - |r|^2$. The reflection coefficient $r$ is obtained as

$$r = -e^{-2ik_{z,3}\tau} R_+/R_-, \quad (4)$$

where we have defined

$$R_\pm = \pm 2i \sin(k_{z,2}\tau)\varepsilon_{x,3}k_{z,2}\omega + 2 \cos(k_{z,2}\tau)\varepsilon_{x,2}k_{z,3}\omega - 2i \sin(k_{z,2}\tau)\sigma k_{z,2}k_{z,3}. \quad (5)$$

The coefficients for the other waves are given as: $A_2 = 2\varepsilon_{x,3}e^{-ik_{z,3}\tau}\omega k_{z,2}/R_-$, and $B_2 = -A_2$. We are now interested to see how the presence of the graphene layer modifies the absorption properties of the metamaterial, and in particular determine the influence of the gate-controlled Fermi level which is tunable *in situ*. The analytical expression for the reflection coefficient *r* allows us to estimate which order of magnitude $E_F$ should have in order to have any influence on the results. If the terms proportional to the conductivity σ are to be comparable in magnitude with the other terms, we see that e.g. $\varepsilon_{x,3}\omega \sim k_{z,3}\sigma$ needs to be satisfied. Since we are considering an EM wave incident from air, $\varepsilon_{x,3} = \varepsilon_{z,3} = \varepsilon_0$. Inserting the expression for σ, one finds that the above equation corresponds to a Fermi level in graphene of

$$E_F \sim \varepsilon_0 c \pi \hbar^2 \sqrt{\omega^2 + t_{\rm rel}^{-2}}/e^2 \; \cos\theta. \tag{6}$$

For grazing incidence $\theta \sim \pi/2$, the graphene layer thus has no influence for any realistic values of $E_F$. We also note that when $\sigma \to \infty$, $r \to 1$ such that the absorption vanishes, $\mathcal{A} \to 0$. In order to gain information about the spatial distribution of the power flow in the structure, we also compute the Poynting vector $\mathbf{S} = \mathbf{E} \times \mathbf{H}$. We write $\mathbf{E} = \mathbf{E}_0 e^{-i\omega t}$ and $\mathbf{H} = \mathbf{H}_0 e^{-i\omega t}$. For future use, we define the normalization constant for the power flow as the time-averaged incident power at $z = \tau$: $S_0 = \varepsilon_0 c/(2\cos\theta)$.

By appropriately tuning $E_F$, graphene-dielectric stacks can exhibit a wavevector dispersion that traces out a hyperbolic contour at a given frequency[21,23]. Such hyperbolic metamaterial (HMM) systems are described by a permittivity tensor whose principle components are opposite in sign[49], and allow for the propagation of evanescent modes in subwavelength imaging[29], and can be designed to exhibit substantial near-field absorption[22]. Such structures are often nonresonant and possess low loss[50]. There have been a variety of HMM systems that have been fabricated[52–54], including semiconductor hybrids, metallic layers[52], and silver nanowires[54]. Gate tunable HMM systems have also been made that allow control of their optical response[53]. Here, we will determine how the absorption properties of HMM structures are modified by the presence of a graphene layer over a broad range of angles θ and Fermi energies $E_F$. Specifically, we focus on gate-tunable HMM structures in the ENZ limit, whereby the tensor components $\varepsilon_{x,2}$ and $\varepsilon_{z,2}$ are opposite in sign. We set $\mu_{y,2} = \mu_0$, and the relaxation time in graphene to $\tau = 10^{-13}$ s. For incident light in the THz regime, with $E_F = 100$ meV and $\lambda = 1.6$ mm, this leads to a surface conductivity of graphene $\sigma = (1.2 \times 10^{-3} + 1.37i \times 10^{-4})$ A·C/J.

We first consider the standard Drude-like frequency response in the metamaterial, such that

$$\varepsilon_{z,2}/\varepsilon_0 = 1 - \alpha^2/[1 + (\alpha f)^2] + i\alpha^3 f/[(1 + (\alpha f)^2]. \tag{7}$$

Here, $\alpha = \lambda/\lambda_z$ and $f = 0.02$, where $\lambda_z$ is the characteristic wavelength, with $\lambda_z = 1.6$ mm. Designing a metamaterial within effective medium theory to possess the prescribed dielectric response in the THz regime can be challenging with current fabrication and materials capabilities. It may be possible however to design a multilayer dielectric structure or rod lattice[55] with polaritonic components to achieve the desired HMM response since the permittivity of polaritonic materials mimic metals in the optical regime[56,57]. Therefore construction of an anisotropic HMM that has a longitudinal ENZ response should be possible at THz frequencies.

It will also be of interest to see how the absorption properties depend on the sign of $\varepsilon_{z,2}$, determined by the frequency of the incident wave, and also its magnitude. We set $\lambda \simeq \lambda_z$, so that $\varepsilon_{z,2} \simeq 0$. If we take the limit $\varepsilon_{x,2} \to 0$, the reflection coefficient simplifies to

$$r = \frac{\varepsilon_0 c - \sigma \cos\theta}{\varepsilon_0 c + \sigma \cos\theta} e^{-2ik_{z,3}\tau}. \tag{8}$$

In the absence of graphene ($\sigma = 0$), we have zero absorption since $|r|^2 = 1$. However, when graphene is present ($\sigma \neq 0$) an interesting opportunity arises. It is seen from the above equation that $r = 0$, meaning *perfect absorption* $\mathcal{A} = 1$ when the following condition is satisfied:

$$\varepsilon_0 c = \sigma \cos\theta. \tag{9}$$

In the low THz regime and below ($\omega \ll 10$ THz), we have $\omega \ll \tau$ and full absorption takes place for angles $\theta_\mathcal{A}$ satisfying

$$\cos\theta_\mathcal{A} = \frac{223.67}{E_F(\rm meV)}. \tag{10}$$

Although this equation represents the criterium for 100% absorption, we show in the next section that the absorption in fact is close to 100% for an extremely wide range of incident angles.

### Results and Discussion

To establish how perfect absorption can be achieved, we discuss how the incident beam can couple to the intrinsic EM modes of the system. Our approach involves finding the poles of the reflection coefficient, Eq. (8), by setting the denominator equal to zero. To find the complete set of poles including the bound and leaky wave contributions, it becomes necessary to extend $k_x = \beta + i\alpha$ to the complex plane and solve the pole dispersion equation (see the Methods section) for β and α (at a given ω). This dispersion equation admits four types of solutions for the complex wavevector[58]. Of these solutions, only those which correspond to perfect absorption modes are retained, since not all solutions correspond to coupling of the incident beam with the structure. The perfect absorption modes of interest are those which are activated via phase-matching of the incident plane wave to solutions corresponding to $k_x/k_0 < 1$, and $\Im m(k_x) = 0$. These modes represent a coherent superposition of waves that propagate without loss along the surface. Indeed, only when the incident wave is incident at the perfect absorption angle, $\theta_\mathcal{A}$, does the direction of energy flow in the air region, $\theta_\mathcal{S}$ ($\equiv \arctan(S_x/S_z)$, precisely coincide with the incident angle $\theta_\mathcal{A}$.

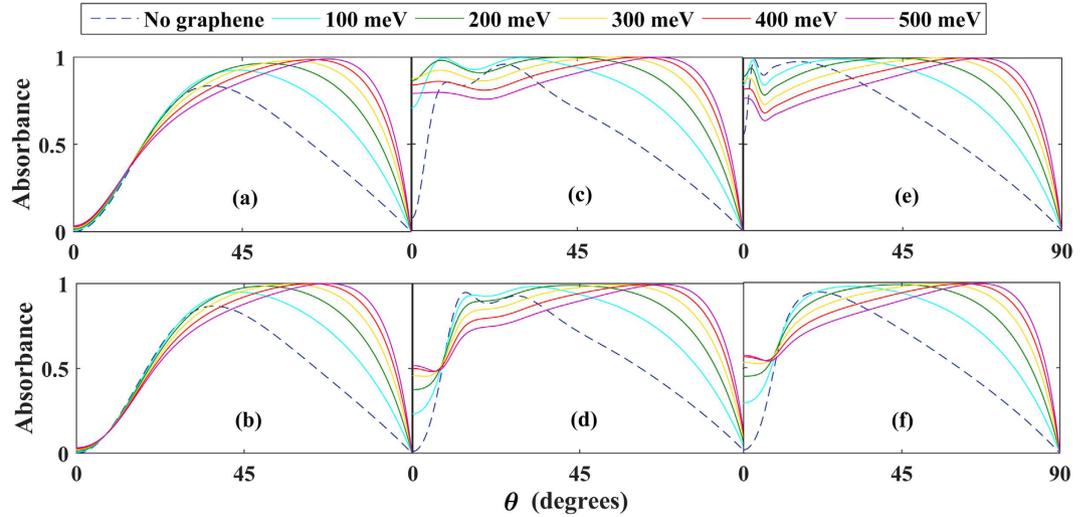

**Figure 2.** Absorption vs. angle of incidence for the Type I case (top row) and Type II case (bottom row) for different thicknesses $\tau$ of the metamaterial: $\tau/\lambda = 0.01$ in (**a,b**), $\tau/\lambda = 0.1$ in (**c,d**), $\tau/\lambda = 0.85$ in (**e,f**). The different curves are obtained for different values of $E_F$. Type I case: $\lambda/\lambda_z = 1.01$ with $\varepsilon_{x,2} = (4 + 0.1i)\varepsilon_0$. Type II case: $\lambda/\lambda_z = 0.99$ and $\varepsilon_{x,2} = (-4 + 0.1i)\varepsilon_0$. In both cases, $\mathrm{Re}\{\varepsilon_{z,2}\} \simeq 0$ and we set the characteristic wavelength of the metamaterial dispersion to $\lambda_z = 1.6$ mm.

Deviations from $\theta_A$ introduce reflected waves which shift the net energy flow in the air region. Additionally, for $r$ to be single-valued, the appropriate branch cuts at the points $\pm k_0$ need to be made for $k_{z,3}$. When examining the perfect absorption phenomena, we utilize a simple yet useful property of the reflection coefficient: Eq. (8) reveals that interchanging the direction of the incident and reflected waves, transforms $r$ into its reciprocal:

$$r(k_{z,3}) = r^{-1}(-k_{z,3}), \tag{11}$$

with $k_{z,3} = k_0 \cos\theta$. This property simplifies the solution process since finding a complex pole, related to $-k_{z,3}$, is equivalent to finding the perfect absorption modes for $k_{z,3}$[59,60]. We underline that the absorption occurring in this way is an interference phenomenon and not based on losses.

We now introduce more possibilities for enhanced absorption, and investigate anisotropic systems where the permittivity components perpendicular and parallel to the interfaces are of opposite sign. The corresponding HMM dispersion relations are then identified as either type I if $\varepsilon_{x,2} > 0$ and $\varepsilon_{z,2} < 0$, or type II if $\varepsilon_{x,2} < 0$ and $\varepsilon_{z,2} > 0$. The operating wavelength of the incident beam is tuned around $\lambda_z$ to control the sign of $\varepsilon_{z,2}$. The component of $\varepsilon$ parallel to the interface always has the same magnitude, but can differ in sign according to $\varepsilon_{x,2} = (\pm 4 + 0.1i)\varepsilon_0$.

Extremely wide-angle absorption for type-I and type-II anisotropic structures is demonstrated clearly in Fig. 2, where we provide results for these two cases for different thicknesses of the metamaterial and values of the Fermi level. The difference between Type I and Type II for a fixed thickness primarily occurs around normal incidence, $\theta = 0$, where the absorption is much higher in the Type I case. It is seen that when the thickness is close to the wavelength [$\tau/\lambda = 0.85$ in panels (e) and (f)], the presence of graphene provides very high absorption for a remarkably wide range of angles. This should be contrasted to the case without graphene (dashed line), where the absorption peaks for an angle $\theta < 45°$ and then rapidly declines. For sufficiently small metamaterial thicknesses, and smaller angles, the metamaterial layer becomes insignificant and the EM wave interacts with mainly the graphene sheet and metal backing. Tuning $E_F$ is seen to improve the absorption dramatically not only in magnitude, but also in terms of which angles of incidence that become absorbed. *Remarkably, the absorption coefficient can be tuned to be nearly perfect, for an unusually broad range of angles.* This is distinct from a scenario involving conventional dielectrics where isolated absorption peaks can appear at specific angles of incidence.

To pinpoint the mechanism of the observed perfect absorption, the mode characteristics of the system should be identified. By correlating the longitudinal wavevector component $k_x$ of the incident EM wave with the permitted EM modes of the graphene-based ENZ structure, the critical coupling responsible for perfect absorption can be identified. Since $k_x/k_0 < 1$, and $\mathfrak{Im}(k_x) = 0$, we search for the allowed fast-wave EM modes (whereby the phase velocities exceed $c$) that the structure supports[51]. By solving the transcendental equation $R_+ = 0$ [Eq. (5)], we find the permitted incident wavevector components that yield $\mathcal{A} = 0$. The obtained $k_x$ are converted to a critical coupling angle via $\theta = \arcsin(k_x/k_0)$, giving the necessary incident beam angles for fast-wave mode coupling to occur. To illustrate this, in Fig. 3 the perfect absorption curves (solutions to $R_+ = 0$) are shown as a function of $E_F$ and $\theta$. The wavelength considered $\lambda = 1.01\lambda_z$ corresponds to a negative real part of $\varepsilon_{z,2}$ that is also near zero. In (a) $\varepsilon_{x,2} = (4 + 0.1i)\varepsilon_0$ corresponding to a type I HMM, while in (b) both permittivity components are of the same sign, since now $\varepsilon_{x,2} = (-4 + 0.1i)\varepsilon_0$. The dispersion curves are shown together with density plots of the absorption $\mathcal{A}$, which gives a global view of the high absorption regions spanned by $E_F$ and $\theta$. Clearly the fast-wave dispersion curves overlap with the regions of perfect absorption. A broader range of incident angles leading to perfect absorption is found in the graphene/HMM structure (a) compared to (b) where both $\mathfrak{Re}\{\varepsilon_{x,2}\}$ and $\mathfrak{Re}\{\varepsilon_{z,2}\}$ are of the same sign. It is evident that incident angles corresponding to perfect absorption occur when a fast-wave mode is excited in the structure, due to the coherent superposition of waves with $k_x$ below the light line. These modes are distinct from conventional guided wave modes that have $k_x/k_0 > 1$, and consequently the EM fields would be evanescent in the air region near the graphene layer. They also differ from leaky wave modes[58] that dissipate when propagating along $x$ since in that case, $\Im m\{k_x\} \neq 0$. Therefore, by coupling the incident beam to fast wave modes of the structure, the enhanced absorption features provided by the HMM in the ENZ regime[61], complimented with the tunable $E_F$ afforded by graphene, generates a broad range of angles for perfect absorption to arise.

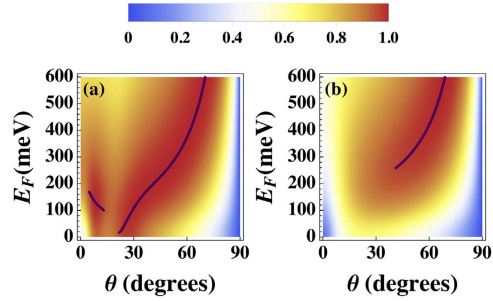

**Figure 3. Absorption as a function of $E_F$ and angle of incidence θ.** The anisotropic metamaterials each have a thicknesses of $\tau = 0.1\lambda$, and the incident beam has wavelength $\lambda = 1.01\lambda_z$, corresponding to $\Re\{\varepsilon_{2,z}\} < 0$. In (**a**) $\varepsilon_{x,2} = (4 + 0.1i)\varepsilon_0$ corresponding to a type I HMM, and in (**b**) $\varepsilon_{x,2} = (-4 + 0.1i)\varepsilon_0$. The fast-wave solutions to the dispersion relation $R_-(-k_{z3}) = 0$ in Eq. (5) are shown by the bold curves.

The behavior of the energy flow is exhibited in Fig. 4, where the Poynting vector components $S_x$ and $S_z$ are shown within the air ($z/\tau > 1$) and metamaterial ($z/\tau \leq 1$) regions for the structure considered in Fig. 3(a). It is evident that magnitudes of $S_x$ and $S_z$ are largest near the interfaces, except for near normal angles of incidence, whereby the energy flows along the metal backing. The flow of $S_x$ in the metamaterial is opposite to the incident beam, demonstrating negative refraction that arises from the small negative permittivity component $\varepsilon_{z,2}$. The perfect absorption features seen in Fig. 3(a) are also consistent with Fig. (d) for $\theta = 41°$ where the energy flow is maximal in the vacuum region at this angle, indicating purely downward energy flow with no upward contributions from surface reflections.

## Methods

We here provide some details for how to obtain the solution of the Maxwell equations in the presence of graphene. The wave-equations for $E_x$ and $E_z$ take the same form in region $j = 2, 3$:

$$\frac{\partial^2 E_k}{\partial z^2} + E_k(\omega^2 \varepsilon_{x,j}\mu_{y,j} - \varepsilon_{x,j}k_x^2/\varepsilon_{z,j}) = 0, \quad k = x, z. \tag{12}$$

The dispersion relation then reads: $\hat{k}_z^2 = \varepsilon_x \mu_y - \varepsilon_x \hat{k}_x^2/\varepsilon_z$, where we have defined $\hat{k}_{x,z} = k_{x,z}/\omega$. An electric field propagating in region $j$ is then written: $\mathbf{E}_j = E_{x,j}^0 e^{ik_x x + ik_{z,j} z}\hat{x} + E_{z,j}^0 e^{ik_x x + ik_{z,j} z}\hat{z}$. The amplitudes are related via: $E_{z,j}^0 = -E_{x,j}^0 \varepsilon_{x,j} k_x/(\varepsilon_{z,j} k_{z,j})$. Using this relation, we may now write down the solution of the electric field in region $j$ which takes into account both transmitted and reflected waves:

$$\mathbf{E}_j = \left[A_j(\hat{x} + \hat{z}\kappa_j)e^{-ik_{z,j}z} + B_j(\hat{x} - \hat{z}\kappa_j)e^{ik_{z,j}z}\right]e^{ik_x x}, \tag{13}$$

where we defined $\kappa_j = k_x \varepsilon_{x,j}/k_{z,j}\varepsilon_{z,j}$. Since the light is incident from air, we take $A_3 = 1$. The magnetic field is obtained as $\mathbf{H}_j = H_{y,j}\hat{y}$ where:

$$H_{y,j} = \frac{\varepsilon_{x,j}}{\hat{k}_{z,j}}(B_j e^{ik_{z,j}z} - A_j e^{-ik_{z,j}z})e^{ik_x x}. \tag{14}$$

There are three coefficients to be determined: $\{A_2, B_2, B_3\}$. To do this, we need to specify the boundary conditions at the air/anisotropic medium interface ($z = \tau$) and the anisotropic medium/PEC interface ($z = 0$). Since the tangential component of $\mathbf{E}$ is always continuous, it follows that $\mathbf{E}_2 \cdot \hat{x}\big|_{z=\tau} = \mathbf{E}_3 \cdot \hat{x}\big|_{z=\tau}$, and $\mathbf{E}_2 \cdot \hat{x}\big|_{z=0} = 0$. The presence of the graphene layer now enters in the boundary condition for the tangential component of $\mathbf{H}$ by writing the free current as $\mathbf{J}_f = \sigma\mathbf{E}$: $H_{y,2}\big|_{z=\tau} = H_{y,3}\big|_{z=\tau} + \sigma(\mathbf{E}_2 \cdot \hat{x})\big|_{z=\tau}$. Specifically, we find that the boundary conditions may be written in the form $\overline{M}\mathbf{a} = \mathbf{b}$ where

$$\mathbf{a} = (A_2, B_2, B_3), \quad \mathbf{b} = (0, e^{-ik_{z,3}\tau}, -\varepsilon_{x,3}e^{-ik_{z,3}\tau}/\hat{k}_{z,3}),$$

$$M = \begin{pmatrix} 1 & 1 & 0 \\ e^{-ik_{z,2}\tau} & e^{ik_{z,2}\tau} & -e^{ik_{z,3}\tau} \\ -e^{-ik_{z,2}\tau}\zeta_+ & e^{ik_{z,2}\tau}\zeta_- & -e^{ik_{z,3}\tau}\varepsilon_{x,3}/\hat{k}_{z,3} \end{pmatrix}, \tag{15}$$

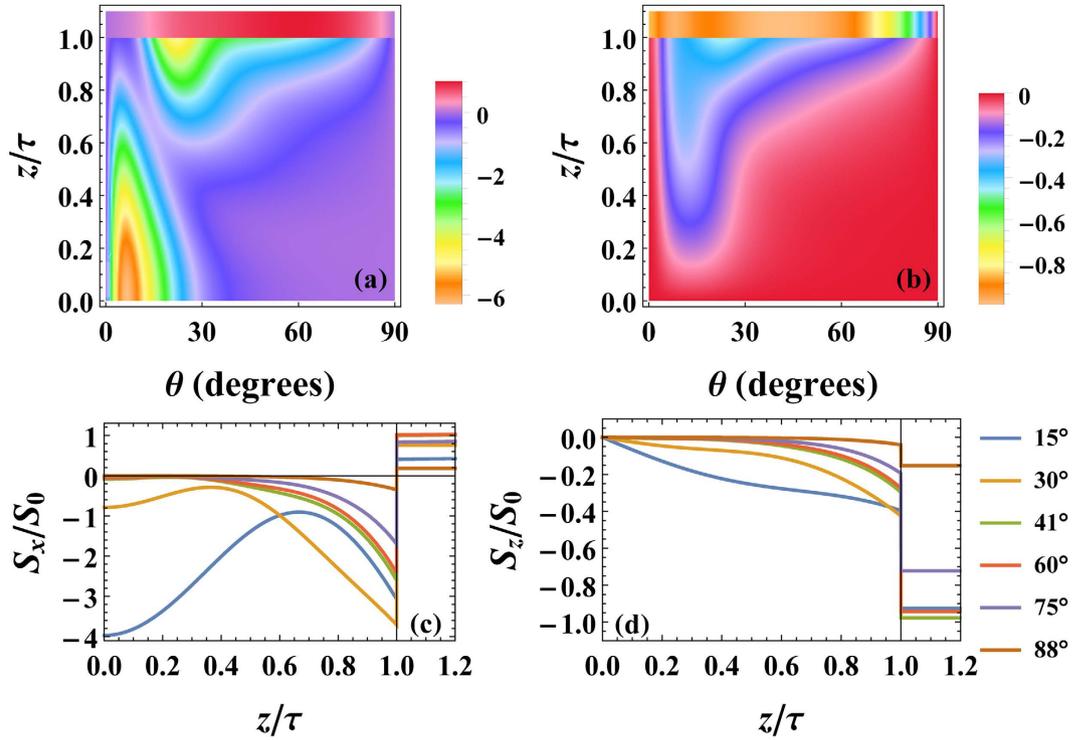

**Figure 4.** Components of the normalized Poynting vector: (**a**) $S_x/S_0$ (along the interface) and (**b**) $S_z/S_0$ (normal to the interface) as functions of position $z/\tau$ and angle θ. The Fermi energy is set at $E_F = 200$ meV. The bottom panels reveal the spatial behavior of $S_x$ and $S_z$ for several θ. Vertical lines identify the vacuum/metamaterial interface at $z/\tau = 1$. The system parameters are the same as those used in Fig. 3(a).

where we defined $\zeta_\pm = \varepsilon_{x,2}/\hat{k}_{z,2} \pm \sigma$. Thus, the coefficients are solved by $\mathbf{a} = M^{-1}\mathbf{b}$, which in turn fully determines $\mathbf{E}$ and $\mathbf{H}$ throughout the entire structure.

We also provide an explicit expression for the time-averaged Poynting vector $\langle \mathbf{S} \rangle = \frac{1}{2} \text{Re}\{\mathbf{E} \times \mathbf{H}^*\}$ in the metamaterial region 2 is:

$$\langle S \rangle_x = \frac{1}{2} \frac{\hat{k}_x |\varepsilon_{x,j}|^2}{|\hat{k}_{z,j}|^2} \text{Re}\{\varepsilon_{z,j}^{-1}(|A_j|^2 e^{2\,\text{Im}\{k_{z,j}\}z} + |B_j|^2 e^{-2\,\text{Im}\{k_{z,j}\}z} - 2\,\text{Re}\{A_j^* B_j e^{2i\text{Re}\{k_{z,j}\}z}\})\},$$

$$\langle S \rangle_z = \frac{1}{2} \text{Re}\left\{\frac{\varepsilon_{x,j}^*}{\hat{k}_{z,j}^*}(|B_j|^2 e^{-2\,\text{Im}\{k_{z,j}\}z} - |A_j|^2 e^{2\,\text{Im}\{k_{z,j}\}z} - 2i\text{Im}\{A_j^* B_j e^{2i\text{Re}\{k_{z,j}\}z}\})\right\}.$$

(16)

To determine the fast-wave modes of the system, we solve Maxwell's equations for the EM fields in the absence of an incident field, thereby effectively treating the structure as a waveguide. This is equivalent to finding the poles of the reflection coefficient for a given sign of $k_{z,3}$. The corresponding transcendental equation determines the allowed ω and $k_x$ pair that a source field can excite in the metamaterial system. From the many possible solutions, we consider only fast-waves corresponding to $k_x < k_0$, and $\Im\{k_x\} = 0$. The dispersion equation that must be solved is,

$$-\varepsilon_0 + \frac{i\varepsilon_{x,2}k_{z,3}}{k_{z,2}} \cot(k_{z,2}\tau) + \frac{\sigma k_{z,3}}{\omega} = 0,$$

(17)

where $k_{z,3}$ can be either positive or negative. When the system under consideration is isotropic, the transcendental equation for the modes in the ENZ limit ($\varepsilon_{x,2} = \varepsilon_{z,2} \to 0$), reduces to,

$$-\varepsilon_0 + \frac{\sigma k_{z,3}}{\omega} = 0.$$

(18)

The $k_x$ is converted into a coupling angle $\theta_c$ that the incident beam must be directed at in order for fast-wave modes to be excited: $\theta_c = \arcsin(k_x/k_0)$. What constitutes a zero numerically in the root finding algorithm that solves Eq. (17) is a tolerance that yields absorption with $\mathcal{A} > 99\%$.

For completeness, we also briefly show results for the spectral dependence of the absorbance, i.e. when moving away from the ENZ regime by varying the wavelength λ of the incident wave. This is shown in Fig. 5 for different thicknesses of the metamaterial layer. As seen, the broad and high absorbance only appears when $\lambda/\lambda_z$ is relatively close to 1. For large deviations from $\lambda/\lambda_z = 1$, a conventional scenario with isolated absorption peaks arises.

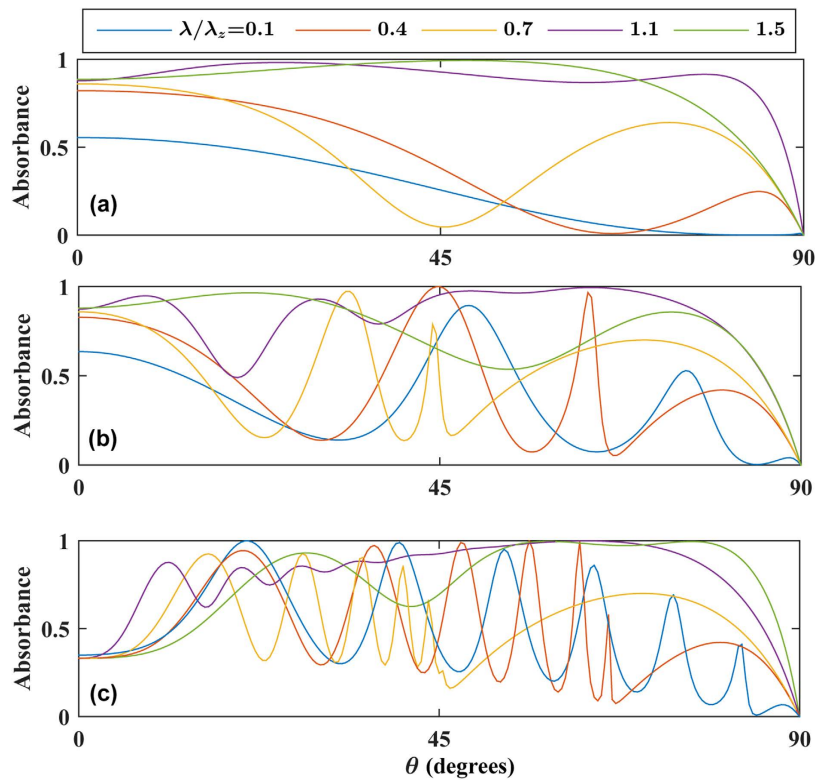

**Figure 5. Absorbance vs. angle of incidence for different wavelengths of the incident wave, in order to show the behavior as one moves away from the ENZ regime.** The thickness of the metamaterial is set to (**a**) $\tau/\lambda = 0.01$, (**b**) $\tau/\lambda = 0.6$, (**c**) $\tau/\lambda = 1.5$. We have fixed $E_F = 300$ meV, $\varepsilon_{x,2} = (4 + 0.1i)\varepsilon_0$, and $\lambda_z = 1.6$ mm.

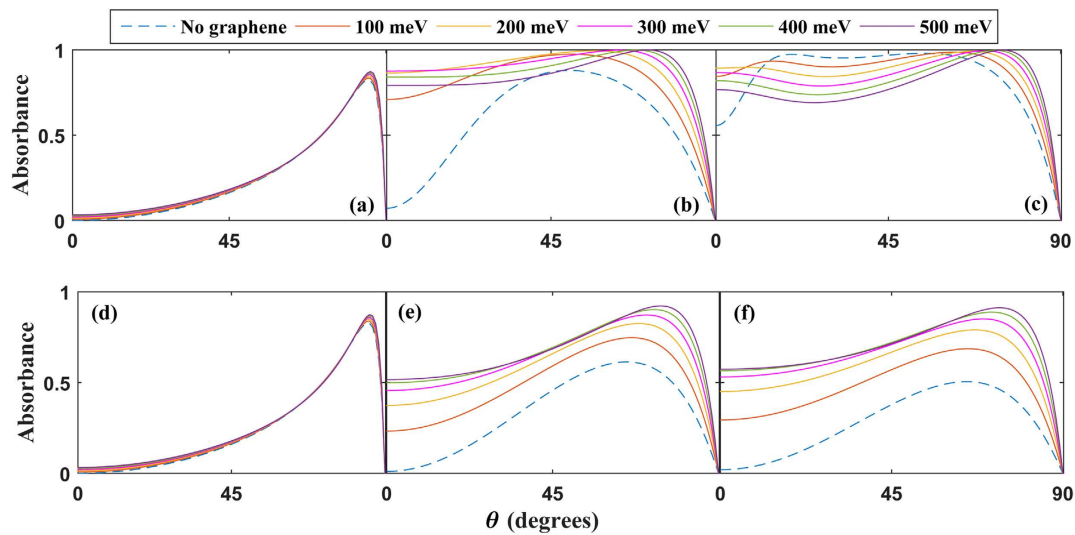

**Figure 6. Absorption vs. angle of incidence for the Type I case (top row) and Type II case (bottom row).** Here we consider the high loss regime where $\mathrm{Im}\{\varepsilon_{z,2}\}/\varepsilon_0 = 1$, while all other parameters are the same as those used in Fig. 2.

We also examine the effects of loss on the absorption properties of the graphene-based metamaterial. In Fig. 6, we consider the same system that was studied in Fig. 2 except now substantial losses are included whereby the component of the permittivity orthogonal to the interface, $\varepsilon_{z,2}$, has a large imaginary component. Although the wide-angle absorption properties are seen to be reduced in the type-II case, the top panels (b) and (c) reveal that remarkably for the type-I HMM, wide-angle absorption remains intact when considerable losses are present.

## References

1. M. Silveirinha & N. Engheta. Tunneling of Electromagnetic Energy through Subwavelength Channels and Bends using ε-Near-Zero Materials. *Phys. Rev. Lett.* **97,** 157403 (2006).
2. J. S. Marcos, M. G. Silveirinha & N. Engheta. μ-near-zero supercoupling. *Phys. Rev. B* **91,** 195112 (2015).
3. R. W. Ziolkowski. Propagation in and scattering from a matched metamaterial having a zero index of refraction. *Phys Rev. E* **70,** 046608 (2004).
4. A. Alù, M. Silveirinha, A. Salandrino & N. Engheta. Epsilon-near zero metamaterials and electromagnetic sources: Tailoring the radiation phase pattern. *Phys. Rev. B* **75,** 155410 (2007).
5. M. Silveirinha & N. Engheta. Design of matched zero-index metamaterials using nonmagnetic inclusions in epsilon-near-zero media. *Phys. Rev. B* **75,** 075119 (2007).
6. R. J. Pollard *et al.* Optical Nonlocalities and Additional Waves in Epsilon-Near-Zero Metamaterials. *Phys. Rev. Lett.* **102,** 127405 (2009).
7. L. V. Alekseyev *et al.* Uniaxial epsilon-near-zero metamaterial for angular filtering and polarization control. *Appl. Phys. Lett.* **97,** 131107 (2010).
8. R. Maas, J. Parsons, N. Engheta & A. Polman. Experimental realization of an epsilon-near-zero metamaterial at visible wavelengths. *Nature Photon.* **7,** 907 (2013).
9. J. Gao *et al.* Experimental realization of epsilon-near-zero metamaterial slabs with metal-dielectric multilayers. *Appl. Phys. Lett.* **103,** 051111 (2013).
10. X. Yang *et al.* Experimental demonstration of near-infrared epsilon-near-zero multilayer metamaterial slabs. *Opt. Express* **21,** 196105 (2013).
11. E. Sachet *et al.* Dysprosium-doped cadmium oxide as a gateway material for mid-infrared plasmonics. *Nat. Mat.* **14,** 414 (2015).
12. J. Kim *et al.* Optical Properties of Gallium-Doped Zinc Oxide-A Low-Loss Plasmonic Material: First-Principles Theory and Experiment. *Phys. Rev. X* **3,** 041037 (2013).
13. J. Kim *et al.* Role of epsilon-near-zero substrates in the optical response of plasmonic antennas. *Optica* **3** 339 (2016).
14. I. I. Smolyaninov & V. N. Smolyaninova. Metamaterial superconductors. *Phys. Rev. B* **91,** 094501 (2015).
15. V. N. Smolyaninova *et al.* Using metamaterial nanoengineering to triple the superconducting critical temperature of bulk aluminum. *Sci. Reports* **5,** 15777 (2015).
16. S. Campione & F. Capolino. Composite material made of plasmonic nanoshells with quantum dot cores: loss-compensation and ε-near-zero physical properties. *Nanotech.* **23,** 235703 (2012).
17. I. B. Vendik *et al.* Tunable Metamaterials for Controlling THz Radiation. *IEEE Trans. Terahertz Sci. Technol.* **2,** 538 (2012).
18. S. H. Lee *et al.* Switching terahertz waves with gate-controlled active graphene metamaterials. *Nat. Mat.* **11,** 936 (2012).
19. Y. V. Bludov, N. M. R. Peres & M. I. Vasilevskiy. Unusual reflection of electromagnetic radiation from a stack of graphene layers at oblique incidence. *J. Opt.* **15,** 114004 (2013).
20. K. V. Sreekanth, A. De Luca & G. Strangi. Negative refraction in graphene-based hyperbolic metamaterials. *Appl. Phys. Lett.* **103,** 023107 (2013).
21. I. V. Iorsh, I. S. Mukhin, I. V. Shadrivov, P. A. Belov & Y. S. Kivshar. Hyperbolic metamaterials based on multilayer graphene structures. *Phys. Rev. B* **87,** 075416 (2013).
22. M. A. K. Othman, C. Guclu & F. Capolino. Graphene-based tunable hyperbolic metamaterials and enhanced near-field absorption. *Opt. Express* **6,** 7614 (2013).
23. M. A. K. Othman, C. Guclu & F. Capolino. Graphene-dielectric composite metamaterials: evolution from elliptic to hyperbolic wavevector dispersion and the transverse epsilon-near-zero condition. *J. Nanophotonics* **7,** 073089 (2013).
24. L. Zhang *et al.* Tunable bulk polaritons of graphene-based hyperbolic metamaterials. *Opt. Express* **22,** 14022 (2014).
25. Y. Xiang *et al.* Critical coupling with graphene-based hyperbolic metamaterials. *Sci. Rep.* **4,** 5483 (2014).
26. A. Andryieuski & A. V. Lavrinenko. Graphene metamaterials based tunable terahertz absorber: effective surface conductivity approach. *Opt. Express* **21,** 9144 (2013).
27. Y. Zhang, Y. Feng, B. Zhu, J. Zhao & T. Jiang. Graphene based tunable metamaterial absorber and polarization modulation in terahertz frequency. *Opt. Express* **22,** 22743 (2014).
28. A. A. Sayem, A. Shahriar, M. R. C. Mahdy & Md. S. Rahman. Control of Reflection through Epsilon near Zero Graphene based Anisotropic Metamaterial. *Proc. ICECE* 812–815 (2014).
29. T. Zhang, L. Chen & X. Li. Graphene-based tunable broadband hyperlens for far-field subdiffraction imaging at mid-infrared frequencies. *Opt. Express* **21,** 20888 (2013).
30. B. D. Xu, C.-Q. Gu, Z. Li & Z.-Y. Niu. A novel structure for tunable terahertz absorber based on graphene. *Opt. Express* **21,** 23803 (2013).
31. A. A. Sayem, M. R. C. Mahdy, I. Jahangir & Md. S. Rahman. Ultrathin Ultra-broadband Electro-Absorption Modulator based on Few-layer Graphene based Anisotropic Metamaterial. *arXiv*:1504.03419 (2015).
32. R. Alaee, M. Farhat, C. Rockstuhl & F. Lederer. A perfect absorber made of a graphene micro-ribbon metamaterial. *Opt. Express* **20,** 28017 (2012).
33. J. Zhang *et al.* Coherent perfect absorption and transparency in a nanostructured graphene film. *Opt. Express* **22,** 12524 (2014).
34. Y. Zou, P. Tassin, T. Koschny & C. M. Soukoulis. Interaction between graphene and metamaterials: split rings vs. wire pairs. *Opt. Express* **20,** 12199 (2012).
35. A. Vakil & N. Engheta. Transformation Optics Using Graphene. *Science* 10, **332,** 1291 (2011).
36. N. Papasimakis *et al.* Graphene in a photonic metamaterial. *Opt. Express* **18,** 8353 (2010).
37. Y. Zhou *et al.* Terahertz wave reflection impedance matching properties of graphene layers at oblique incidence. *Carbon* **96,** 1129 (2016).
38. Santos, J. E. *et al.* Renormalization of nanoparticle polarizability in the vicinity of a graphene-covered interface. *Phys. Rev. B* **90,** 235420 (2014).
39. Ning, R. *et al.* Dual-gated tunable absorption in graphene-based hyperbolic metamaterial. *Aip Advances* **5,** 067106 (2015).
40. C. S. R. Kaipa *et al.* Enhanced transmission with a graphene-dielectric microstructure at low-terahertz frequencies. *Phys. Rev. B* **85,** 245407 (2012).
41. A. Fallahi & J. P.-Carrier. Design of tunable biperiodic graphene metasurfaces. *Phys. Rev. B* **86,** 195408 (2012).
42. M. Jablan, H. Buljan & M. Soljačić. Plasmonics in graphene at infrared frequencies. *Phys. Rev. B* **80,** 245435 (2009).
43. T. Low & P. Avouris. Graphene Plasmonics for Terahertz to Mid-Infrared Applications. *ACS Nano* **8,** 1086 (2014).
44. Z. Fei *et al.* Gate-tuning of graphene plasmons revealed by infrared nano-imaging. *Nature* **487,** 82 (2012).
45. L. Ju *et al.* Graphene plasmonics for tunable terahertz metamaterials. *Nat. Nanotechnol.* **6,** 630 (2011).
46. K. V. Sreekanth & T. Yu. Long range surface plasmons in a symmetric graphene system with anisotropic dielectrics. *J. Opt.* **15,** 055002 (2013).
47. L. A. Falkovsky & S. S. Pershoguba. Optical far-infrared properties of a graphene monolayer and multilayer. *Phys. Rev. B* **76,** 153410 (2007).
48. M. Liu *et al.* A graphene-based broadband optical modulator. *Nature* **474,** 64 (2011).
49. D. R. Smith & D. Schurig. Electromagnetic wave propagation in media with indefinite permittivity and permeability tensors. *Phys. Rev. Lett.* **90,** 077405 (2003).

50. C. L. Cortes, W. Newman, S. Molesky & Z. Jacob. Quantum nanophotonics using hyperbolic metamaterials. *J. Opt*. **14,** 063001 (2012).
51. S. Feng & K. Halterman. Coherent perfect absorption in epsilon-near-zero metamaterials. *Phys. Rev. B* **86,** 165103 (2012).
52. V. Caligiuri *et al.* Dielectric singularity in hyperbolic metamaterials: the inversion point of coexisting anisotropies. *Sci. Rep*. **6,** 20002 (2016).
53. G. T. Papadakis & H. A. Atwater. Field-effect induced tunability in hyperbolic metamaterials. *Phys. Rev. B* **92,** 184101 (2015).
54. K.-T. Tsai *et al.* Looking into Meta-Atoms of Plasmonic Nanowire Metamaterial. *Nano Lett.* **14,** 4971 (2014).
55. M. Massaouti *et al.* Eutectic epsilon-near-zero metamaterial terahertz waveguides *Opt. Lett.* **38,** 1140 (2013).
56. K. C. Huang *et al.* Field Expulsion and Reconfiguration in Polaritonic Photonic Crystals. *Phys. Rev. Lett.* **90,** 196402 (2003).
57. M. Kafesaki, A. A. Basharin, E. N. Economou & C. M. Soukoulis. THz metamaterials made of phonon-polariton materials. *Photon. Nanostruct.: Fund. Appl.* **12,** 376 (2014).
58. K. Halterman, S. Feng & V. C. Nguyen. Controlled leaky wave radiation from anisotropic epsilon near zero metamaterials. *Phys. Rev. B* **84,** 075162 (2011).
59. M. Neviere & P. Vincent. Brewster phenomena in a lossy waveguide used just under the cut-off thickness. *J. Opt.* **11,** 153 (1980).
60. T. S. Luk *et al.* Directional perfect absorption using deep subwavelength low-permittivity films. *Phys. Rev. B* **90,** 085411 (2014).
61. K. Halterman & J. M. Elson. Near-perfect absorption in epsilon-near-zero structures with hyperbolic dispersion. *Opt. Express* **22,** 7337 (2014).

### Acknowledgements

J.L. acknowledges support from the Outstanding Academic Fellows programme at NTNU and the Norwegian Research Council Grant No. 205591, No. 216700, and No. 240806. K.H. was supported in part by the NAWC Nanomaterials Core S&T Network and a grant from the Department of Defense High Performance Computing Modernization Program.

### Author Contributions

K.H. and J.L. contributed to the analytical and numerical calculations as well as the discussion of the results and the writing of the manuscript.

### Additional Information

**Competing financial interests:** The authors declare no competing financial interests.

**How to cite this article**: Linder, J. and Halterman, K. Graphene-based extremely wide-angle tunable metamaterial absorber. *Sci. Rep.* **6**, 31225; doi: 10.1038/srep31225 (2016).